\documentclass[a4paper]{article}

\usepackage{INTERSPEECH2019}
\usepackage{booktabs}
\usepackage{cmll}
\usepackage{combelow}
\usepackage[usenames,dvipsnames]{xcolor}
\usepackage[inline]{enumitem}
\usepackage{hyperref}
\usepackage{fontawesome}
\usepackage{microtype}
\usepackage{url}
\usepackage{tikz}
\usepackage{xspace}

\usepackage{mymacros}

\hypersetup{
    colorlinks,
    citecolor=Violet,
    linkcolor=Red,
    urlcolor=Blue}

\title{%
    \textsc{Kite}: Automatic speech recognition for unmanned aerial vehicles
}

\name{%
    Dan Onea\cb{t}\u{a}$^{1,2}$,
    Horia Cucu$^{1,2}$%
}

\address{
  $^1$University \textsc{Politehnica} of Bucharest, Romania\\%
  $^2$Autonomous Systems, Bucharest, Romania%
}

\email{%
    dan.oneata@speed.pub.ro,
    horia.cucu@speed.pub.ro
}

\newcommand{\new}[1]{#1}
\newcommand{\uav}{\textsc{Kite}\xspace}
\newcommand{\uavtrain}{\uav train\xspace}
\newcommand{\uavtest}{\uav eval\xspace}

\begin{document}

\maketitle

\begin{abstract}
    This paper addresses the problem of building a speech recognition system attuned to the control of unmanned aerial vehicles (UAVs).
    Even though UAVs are becoming widespread, the task of creating voice interfaces for them is largely unaddressed.
    To this end, we introduce a multi-modal evaluation dataset for UAV control, consisting of spoken commands and associated images,
    which represent the visual context of what the UAV ``sees'' when the pilot utters the command.
    \new{We provide baseline results and address two research directions:
    \ia how robust the language models are, given an incomplete list of commands at train time;
    \ib how to incorporate visual information in the language model.
    We find that recurrent neural networks (RNNs) are a solution to both tasks:
    they can be successfully adapted using a small number of commands and they can be extended to use visual cues.
    Our results show that the image-based RNN outperforms its text-only counterpart even if the command--image training associations are automatically generated and inherently imperfect.}
    The dataset and our code are available at \url{http://kite.speed.pub.ro}.
\end{abstract}

\section{Introduction}

As unmanned aerial vehicles (UAVs) are reaching consumer-level production,
we expect an increasing effort into making them more accessible.
One way to achieve accessibility is by designing interfaces that are easier to operate.
The typical interface for UAVs relies on windows, icons, menus, pointers (WIMP),
but recent research proposes a variety of interfaces, such as gestures \cite{naseer2013iros,peshkova2017pervasive}, gaze \cite{monajjemi2013iros} or speech \cite{supimros2014,landau2017ichri}.
Our work addresses the last category---controlling an UAV by spoken commands
(we assume the utterances are recorded from the pilot's headset and transcribed by a system located on the ground).


We are interested in transcribing a diverse set of commands,
from simpler, movement-related instructions (such as, \textit{turn right} or \textit{move up})
to more elaborate ones, specific to certain operational scenarios (such as, \textit{zoom on the poacher shooting the rhinoceros} or \textit{what type of trees does the truck carry?}).
Being able to accurately recognize a wide range of instructions is a prerequisite in creating systems that can connect language to perception and action. 
The high-level queries we consider are similar to those encountered in
situated language understanding for human-robot interaction \cite{gorniak2007cognitive} or
visual question answering \cite{antol2015iccv}.

\begin{figure}
    \centering
    \def\w{3.0cm}
    \def\h{2.25cm}
    \footnotesize
    \begin{tabular}{cc}
        \multicolumn{2}{c}{scenario: \textit{documenting protest}} \\
        \includegraphics[width=\w,height=\h]{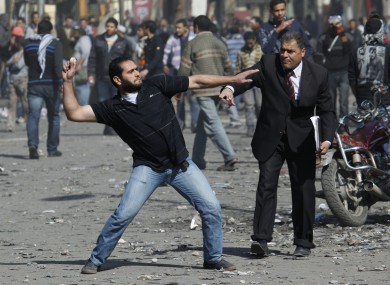} &
        \includegraphics[width=\w,height=\h]{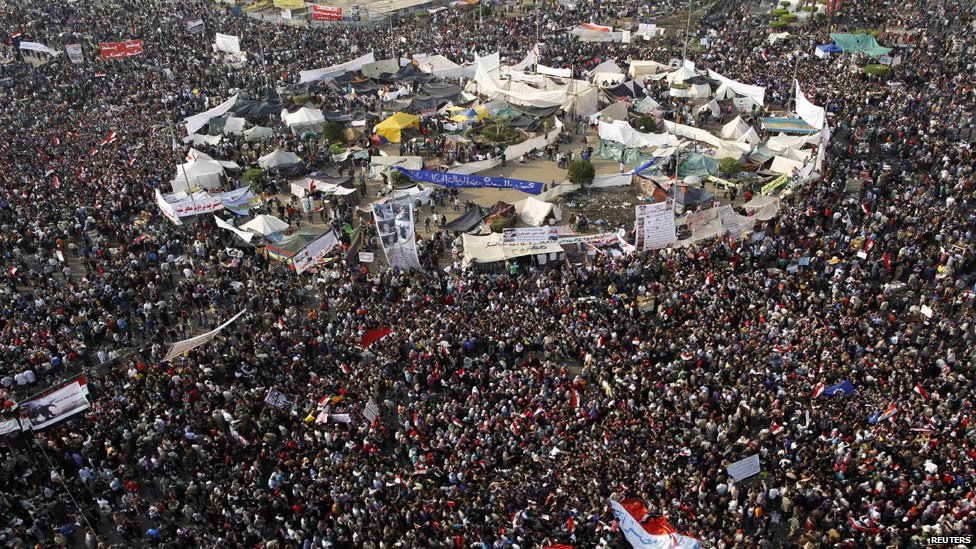} \\
        \faVolumeUp $\,$ \textit{zoom in the} &
        \faVolumeUp $\,$ \textit{estimate the number} \\
        \textit{person throwing stones} &
        \textit{of people in the square} \\
        \midrule
        \multicolumn{2}{c}{scenario: \textit{forest surveillance}} \\
        \includegraphics[width=\w,height=\h]{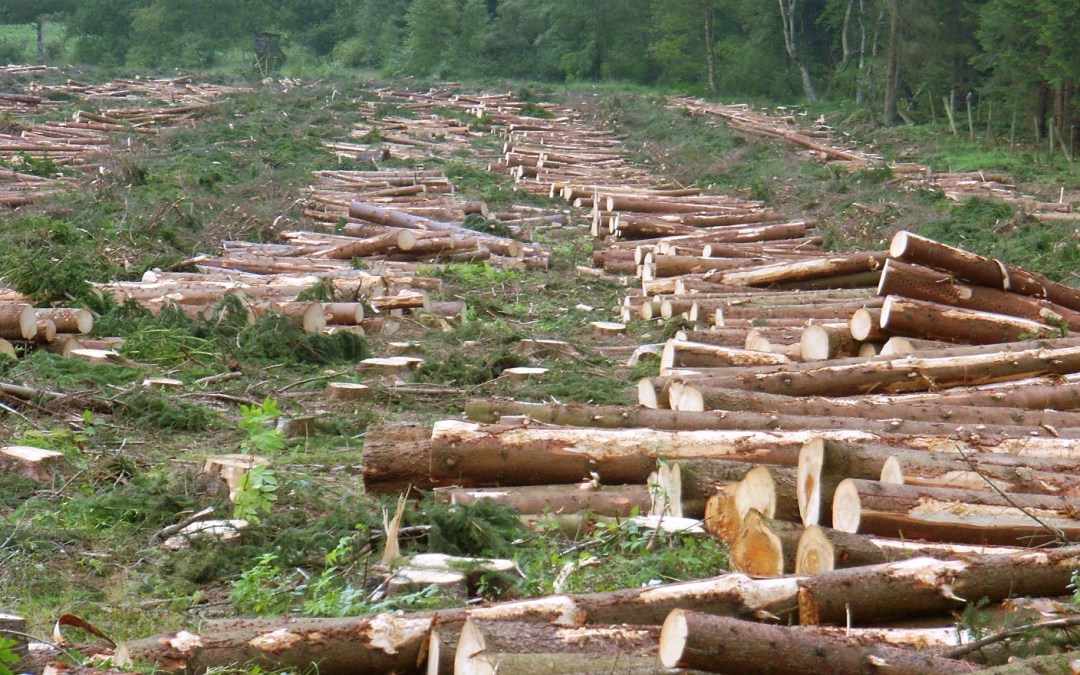} &
        \includegraphics[width=\w,height=\h]{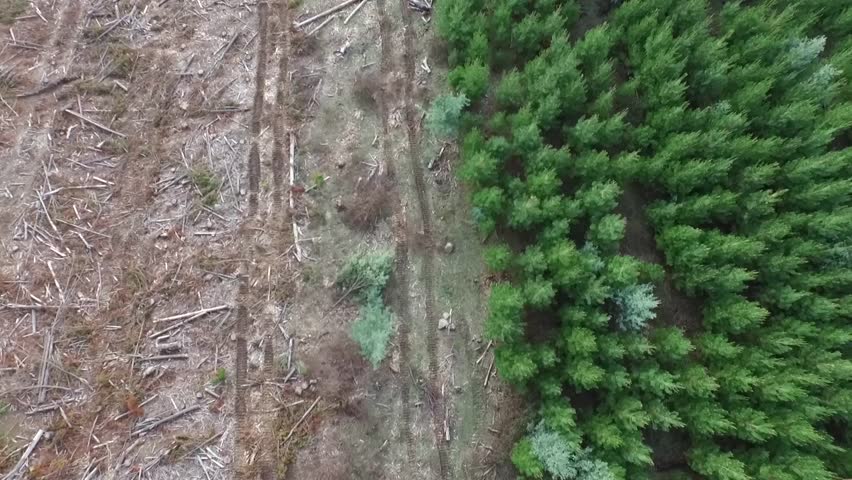} \\
        \faVolumeUp $\,$ \textit{move to the cut trees} &
        \faVolumeUp $\,$ \textit{signal skidding activity} \\
    \end{tabular}
    \caption{%
        Examples of commands and images from \uavtest.
    }
    \label{fig:examples}
\end{figure}

The first step towards building systems for UAV control is having a way of evaluating and comparing them.
In this paper, we propose an evaluation dataset for this task, named \uavtest.
While in other communities (\eg, computer vision) UAV-related datasets are emerging \cite{mueller2016eccv,zhuvisdrone2018arxiv},
we are, to the best of our knowledge, the first to introduce such a database for speech.
In choosing the commands, we took inspiration from UAV pilots and tried to address relevant scenarios in which UAVs could be used;
figure~\ref{fig:examples} shows a sample of commands for two such scenarios.

A baseline method for our task is a generic speech recognition system.
However, since there is a domain mismatch between existing datasets and \uavtest,
we do not expect such a system to perform particularly well.
As an improvement, we consider adapting the generic system to the task at hand;
in particular, we experiment with adapting the language models to UAV commands.
While we can have a good idea of the type of commands given by a pilot in a particular scenario,
it is unavoidable to encounter unforeseen commands at test time.
To quantify such errors, we carry the adaptation procedure by varying the overlap between the training and testing commands.

A given command depends on a number of factors, \eg, visual context, type of scenario, previous commands.
In this paper, we try to leverage the visual information---what the UAV and its pilot are ``seeing'' when a command is uttered.
Consider, for example, an UAV performing a forest surveillance operation, floating close to the border of a forest and a muddy path, as in the bottom right image from figure~\ref{fig:examples};
given this picture, we expect a command like \textit{signal skidding activity} rather than \textit{signal skiing activity}.
To handle such cases,
\new{we modify the language model to incorporate visual information and extend the evaluation set by manually assigning a relevant image to each command in \uavtest.
Collecting manual associations for the training set is an expensive, if not prohibitive, task.}
Instead, we propose a semi-automatic procedure to generate correspondences between images and commands by relying on existing image databases.

To summarize, our main contributions are:
\begin{enumerate*}
    \item We introduce a multi-modal evaluation dataset for UAV control,
        where each command has associated its utterance and a relevant image (\S\ref{sec:dataset}).
    \item We build a baseline speech recognition system by using external data and
        compare it to improved models that are adapted on various amounts of data (\S\ref{sec:methodology} and \S\ref{sec:results}).
    \item We augment the language model to include visual information and
    use semi-automatic procedures to generate command--image associations as training data (\S\ref{sec:methodology} and \S\ref{sec:results}).
\end{enumerate*}

\section{Related work}

We discuss two research directions related to our work.

\textbf{Speech recognition for UAV control.}
The task of speech recognition for UAV control is relatively unexplored
and the few published works on this topic \cite{supimros2014,landau2017ichri,draper2003} focus on recognition of simple commands:
the authors of~\cite{supimros2014} predict a fixed set of nine commands using a classification pipeline based on audio features, such as energy and MFCC;
the method in \cite{draper2003} recognizes commands to navigate through menus, operations which were previously achieved through keyboard presses.

\textbf{Multi-modal learning.}
Systems that use multiple types of sensory data (\eg, audio, visual, language) are known as \textit{multi-modal} systems.
Many works focus on combinations of two out of the three aforementioned modalities.
Arguably, the most common combination is audio--language, as it includes the task of speech recognition,
but the other two combinations, vision--language and audio--visual, are seeing increased attention.

Vision-language systems are used in tasks such as image captioning \cite{kiros2014icml,karpathy2015cvpr} or visual question answering \cite{antol2015iccv,krishna2017ijcv}.
Many such systems model the language in the context of an image:
they estimate the probability distribution over the next word given the preceding words and the visual context.
The most common approach uses a recurrent neural network to model the distribution over the words and a convolutional neural network to extract visual features \cite{karpathy2015cvpr,mao2014arxiv,vinyals2015cvpr,xu2015icml};
we use a similar architecture.
Audio-visual systems target tasks such as
image retrieval by speech \cite{synnaeve2014nips,harwath2016nips,harwath2018eccv},
embedding learning \cite{harwath2016nips,harwath2015asru},
speech-prompted object localization \cite{harwath2018eccv} or
semantic keyword spotting \cite{kamper2019taslp}.
The typical approach exploits statistical correspondences and learns embeddings for the two modalities, utterances and images, to a common sub-space.


The work of Sun \etal \cite{sun2016slt} combines all three modalities and is most similar to ours:
they attempt to improve an ASR system based on a language model that takes the context image as input.
We differ from them by taking other architectural decisions and, more importantly, by assuming a scenario with small amounts of data.
For this reason we have to rely on out-of-domain datasets for initialization and semi-automatic methods to generate training data.

\section{Dataset}
\label{sec:dataset}

In this section we introduce the \textsc{Kite} dataset,
a multi-modal dataset for UAV control.
The dataset consists of three types of modalities:
language (commands),
audio (utterances),
vision (images).
We have build the dataset by first deciding on a set of commands,
then recording the spoken utterances,
and, finally, associating a image to each command.

We identified three types of UAV-specific commands:
\ia movement-related,
\ib camera-related,
and
\ic scenario-specific.
The scenario-specific category was further split by considering seven types of scenarios,
which we thought of interest for UAV applications:
\begin{enumerate*}
  \item \textit{documenting a protest};
  \item \textit{forest surveillance};
  \item \textit{train surveillance};
  \item \textit{anti-poaching operation};
  \item \textit{natural disaster rescue operations};
  \item \textit{ski monitoring};
  \item \textit{sea monitoring}.
\end{enumerate*}
We collaborated with UAV pilots to prepare a list of possible English commands based on these scenarios,
which we then distilled into a finite state grammar (FSG), named \textit{gold FSG}.
The FSG representation has several advantages over a raw list of commands:
\ia it is more compact, given that many commands overlap;
\ib it allows us to sample new commands, which are similar, but not necessarily identical, to the ones proposed by the pilots;
\ic it enables us to create new datasets, which are used in our experimental procedure for training.
%
%

\textbf{Evaluation dataset.}
We selected a set of 2,880 commands and recorded their utterances with the help of 16 L2 English speakers.
Each speaker was assigned 180 commands: 20 movement-related, 20 camera-related, 20 for each of the seven scenarios.
The utterances were recorded using a web application and allowed speakers to use their own recording environment.
On average a command has five words and lasts about 3.5 seconds.
Some examples of commands can be found in figure \ref{fig:examples}.
This dataset is used for evaluation and we name it \uavtest.
The recordings were done in noiseless conditions,
but in order to simulate a real-world scenario we have corrupted the audio files with noise.
We have selected noise samples corresponding to outdoor noises from the MUSAN dataset \cite{snyder2015arxiv}
and added them to the spoken utterances using a signal-noise ratio (SNR) of 10.

\textbf{Training dataset.}
Based on the gold FSG we sampled datasets of different sizes, from 2,048 to 65,536 ($2^{11},\dots,2^{16}$) commands;
the varying number of commands simulates scenarios where we have access to different amounts of data.
In order to account for the variability in the sampling process, we generate five folds for each dataset size.
This dataset is used for training and we name it \uavtrain.
Note that the training set is text-only and is employed in domain adaptation of the language models.
Table \ref{tab:statistics-generated-corpora} reports statistics of the dataset.


\begin{table}
    \centering
    \footnotesize
    \newcommand{\res}[2]{#1 \scriptsize{$\pm$ #2}}
    \caption{%
        Statistics for \uavtrain.
        For each dataset of size $n$, we report the number of unique commands and the number of commands in the evaluation set.
        We report the mean and the standard deviation over the five folds.
        The last row indicates that the FSG can generate over 35K commands.
    }
    \begin{tabular}{rrrr}
        \toprule
        & & \multicolumn{2}{c}{overlap with evaluation set} \\
            \cmidrule(lr){3-4}
        $n$      & unique & number & proportion (\%) \\
        \midrule
        2,048    & \res{ 1,313.2}{12.9}       & \res{1,705.6}{10.1} & \res{59.2}{0.4} \\
        4,096    & \res{ 2,192.4}{37.0}       & \res{1,900.8}{12.3} & \res{66.0}{0.4} \\
        8,192    & \res{ 3,532.2}{22.3}       & \res{2,147.8}{14.1} & \res{74.6}{0.5} \\
        16,384   & \res{ 5,420.0}{52.4}       & \res{2,360.8}{ 8.9} & \res{82.0}{0.3} \\
        32,768   & \res{ 7,795.8}{45.2}       & \res{2,562.0}{ 9.3} & \res{89.0}{0.3} \\
        65,536   & \res{10,587.4}{31.3}       & \res{2,702.4}{ 7.2} & \res{93.8}{0.2} \\
        $\infty$ & 35,753                     & 2,880               & 100 \\
        \bottomrule
    \end{tabular}
    \label{tab:statistics-generated-corpora}
\end{table}

\subsection{Visual extension}

We extended both the training and evaluation parts with a visual component,
by assigning each command with a relevant image (an image that might have been observed when that particular command has been uttered).

\textbf{Evaluation dataset.}
The associations were done by searching images on search engines (such as Google Images) using queries that were related to a given command.
We asked the participants to select images which are taken from a higher perspective, similar to what a UAV would record,
but for some commands it was difficult to meet this requirement.
See figure \ref{fig:examples} for some examples of commands and their associated images.

\textbf{Training dataset.}
Obtaining manual image--command associations for the training data
would have been prohibitively expensive,
so we relied on a semi-automatic approach.
The idea was to link keywords from commands to the image classes from standard computer vision databases.
Here is a example of creating such associations:

\begin{tikzpicture}[
    font=\footnotesize,
    auto,
    thick,
    box1/.style={text width=2.0cm, minimum width=0.5cm, align=center, color=gray!40!black},
    keyword/.style={color=red},
    class/.style={color=blue},
]
    \node[box1] (command 1) at (0,  0) {\textit{watch the street}};
    \node[box1] (command 2) at (0, -1) {\textit{move over the avenue}};
    \node[box1] (command 3) at (0, -2) {\textit{alert on drowning person}};

    \node[keyword] (keyword 1) at (2, -0.5) {\textsf{street}};
    \node[keyword] (keyword 2) at (2, -2.0) {\textsf{marine}};

    \node[class] (class 1) at (3.5,  0) {\textsf{street}};
    \node[class] (class 2) at (3.5, -1) {\textsf{road}};
    \node[class] (class 3) at (3.5, -2) {\textsf{boathouse}};

    \node (image 1) at (5.5,  0) {\includegraphics[width=1.25cm]{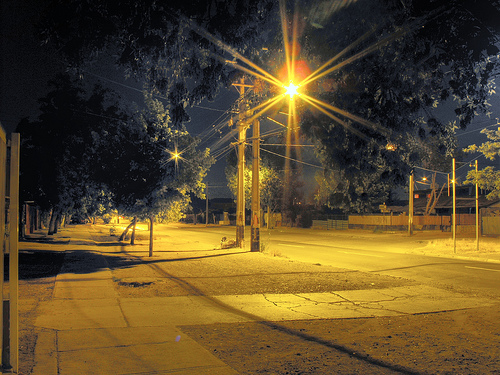}};
    \node (image 2) at (5.5, -1) {\includegraphics[width=1.25cm]{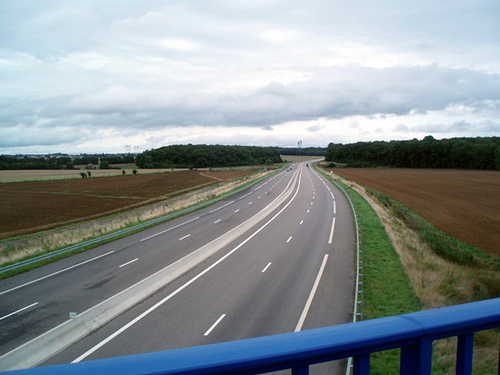}};
    \node (image 3) at (5.5, -2) {\includegraphics[width=1.25cm, height=0.8cm]{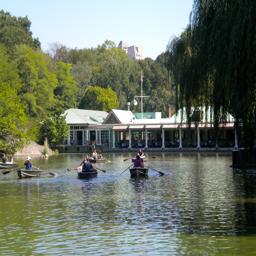}};

    \draw[gray, dashed] (command 1.east) -- (keyword 1);
    \draw[gray, dashed] (command 2.east) -- (keyword 1);
    \draw[gray, dashed] (command 3) -- (keyword 2);

    \draw (keyword 1) -- (class 1.west);
    \draw (keyword 1) -- (class 2.west);
    \draw (keyword 2) -- (class 3);

    \draw[gray, dashed] (class 1) -- (image 1);
    \draw[gray, dashed] (class 2) -- (image 2);
    \draw[gray, dashed] (class 3) -- (image 3);
\end{tikzpicture}

Instead of directly linking each command on the left to a corresponding image on the right,
we just associate keywords (in red) to image classes (in blue).
The effort is much reduced, because there are far fewer keywords and classes than commands and images.
The command--keyword associations are done automatically by searching expressions in the commands,
while the class--image associations are obtained from computer vision datasets, in our case
ImageNet \cite{russakovsky2015ijcv} and MIT Places \cite{zhou2018pami}.

\section{Methodology}
\label{sec:methodology}

\begin{figure}
    \centering
    \begin{tikzpicture}[%
            font=\footnotesize,
            auto,
            thick,
            model/.style={text width=1.25cm, fill=blue!25, minimum height=1.0cm, minimum width=0.5cm, align=center},
            op/.style={text width=1.25cm, fill=yellow!25, minimum height=1.0cm, minimum width=0.5cm, align=center, draw=black},
            dataset/.style={color=gray!70, font=\footnotesize},
            node distance=2.5cm,
        ]
        \node[model] (model acustic)  at (0, 0) {acoustic model};
        \node[model] (model limba 1)  at (2, 0) {language model 1};
        \node[model] (model limba 2)  at (4, 0) {language model 2};

        \node (date acustic) at (0, 1.5) {audio $+$ text};
        \node (date limba 1) at (2, 1.5) {text};
        \node (date limba 2) at (4, 1.5) {text $+$ images$^\star$};

        \node[dataset] (dataset acustic) at (0, 1.8) {TED-LIUM};
        \node[dataset, color=red!70] (dataset limba 1) at (2, 1.8) {\uavtrain};
        \node[dataset] (dataset limba 1) at (2, 2.1) {CANTAB};
        \node[dataset, color=red!70] (dataset limba 2) at (4, 1.8) {\uavtrain$^\star$};
        \node[dataset]               (dataset limba 2) at (4, 2.1) {FlickR-8K$^\star$};
        \node[dataset]               (dataset limba 2) at (4, 2.4) {CANTAB};

        \node[op] (decodare)  at (2, -2.00) {decoding};
        \node[op] (rescore)   at (4, -2.00) {rescoring};
        \node[op] (evaluare)  at (6, -2.00) {evaluation};

        \node (date decodare) at (2, -3.5) {audio};
        \node (date rescore)  at (4, -3.5) {images$^\star$};
        \node (date evaluare) at (6, -3.5) {text};

        \node[dataset, color=red!70] at (2, -3.8) {\uavtest};
        \node[dataset, color=red!70] at (4, -3.8) {\uavtest$^\star$};
        \node[dataset, color=red!70] at (6, -3.8) {\uavtest};

        \draw[gray!50] (-1.25, -1.00) -- (7, -1.00);
        \node[blue!80]   at ( 6.50, -0.75) {\textit{training}};
        \node[orange!80] at (-0.75, -1.25) {\textit{evaluation}};

        \draw[->] (date acustic) edge (model acustic);
        \draw[->] (date limba 1) edge (model limba 1);
        \draw[->] (date limba 2) edge (model limba 2);
        \draw[->] (decodare)     edge (rescore);

        \draw[->] (model acustic) |-   (decodare);
        \draw[->] (model limba 1) edge (decodare);
        \draw[->] (model limba 2) edge (rescore);
        \draw[->] (rescore)       edge (evaluare);
        \draw[->] (date decodare) edge (decodare);
        \draw[->] (date rescore)  edge (rescore);
        \draw[->] (date evaluare) edge (evaluare);
    \end{tikzpicture}
    \caption{%
        Methodological overview.
        Our ASR system consists of an acoustic model and two language models.
        We initialize these components on generic datasets (in gray),
        and then adapt them to domain-specific data (in red).
        The datasets that have a visual component are marked with a star.
    }
    \label{fig:methodology}
\end{figure}
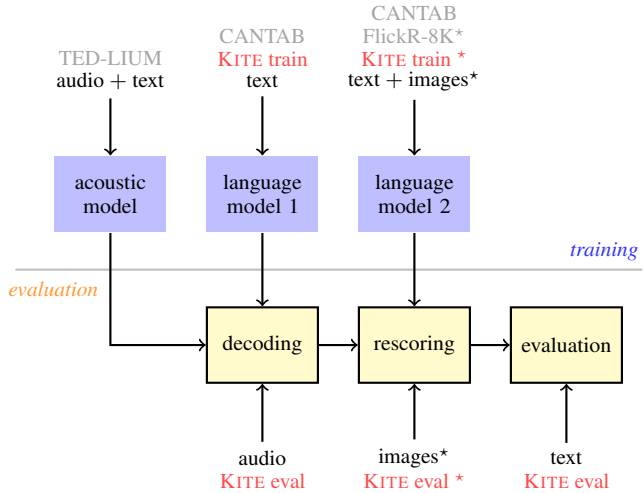

Our speech recognition system is based on an acoustic model and two language models.
The acoustic model consists of a time delay neural network \cite{peddinti2015interspeech} and is implemented in Kaldi \cite{povey2011asru}.
The first language model is used for decoding and it is either a finite state grammar (FSG) or an n-gram.
The second language model is used for re-scoring and, hence, it is richer and more flexible than the first one;
we use either a larger n-gram or a recurrent neural network (RNN).

We make use of existing databases to learn powerful representations.
For acoustics we learn the model entirely on external data (the TED-LIUM dataset),
while for language models we use external datasets to learn an initial set of parameters,
which are then adapted to our task by fine-tuning.
Figure \ref{fig:methodology} shows our systems' components and the use of datasets in our methodology;
these are further presented in subsections~\ref{subsec:language-models} and \ref{subsec:auxiliary-datasets}.

\subsection{Language models}
\label{subsec:language-models}

In the following, we describe the language models and their adaptation procedure to the task of drone control.

\textbf{FSG.}
Given the training corpus of commands, \uavtrain, we use the FSG to construct a language model that allows only the commands that appear in the training set.
Practically, the training commands are joined by the \texttt{or} operator.
The FSG is used only for decoding.
Apart from the FSG built from the training data, we have also experimented with the gold FSG, which was manually created as described in section \ref{sec:dataset}.

\textbf{N-gram.}
We use two types of n-gram models, both of order four, one using 2M n-grams, while the other, 10M.
The smaller n-gram is used for decoding and the larger one for rescoring.
We perform domain adaptation by training a generic model (trained on the CANTAB dataset) and a specific one (trained on \uavtrain)
and then interpolating the two \cite{rosenfeld2000};
we use a coefficient of 0.9 for the domain-specific model.

\textbf{RNN.}
We use a recurrent network with long short-term memory cells (LSTM; \cite{hochreiter1997nc}) to model the probability over the next word given a sequence.
We ``tie'' the input and output embedding matrices \cite{inan2016arxiv,press2017eacl}, as this has been shown to improve performance \cite{merity2017arxiv,melis2018iclr}.
The vocabulary is fixed at 10,000 words and the embedding size and the hidden size of the LSTM cells is set to 512;
the resulting network has around 9.4M parameters.
Inspired by the results of Melis \etal \cite{melis2018iclr}, we optimize the hyper-parameters (\eg, drop-out, learning rate) using an automatic procedure \cite{bergstra2013jmlr}.
We perform domain adaptation by fine-tuning the network \cite{gangireddy2016interspeech}:
we train a generic model on a source dataset (CANTAB),
and then continue updating the weights for a fixed number of epochs (25) on the target dataset (\uavtrain).
The fine-tuning procedure allows us to train on smaller datasets, which otherwise would be difficult to fit.
The RNN language model is used only for rescoring by reordering the n-best list.

\textbf{Multi-modal RNN.}
We extend the RNN with an additional component, an encoder, which takes in an image, extracts features and passes them to the first LSTM cell.
This architecture allows us to estimate the distribution over the next word $w_t$ given the preceding words $\mathbf{w}_{1:t-1}$ and the visual context $\mathbf{v}$, that is, $p(w_t | \mathbf{w}_{1:t-1}, \mathbf{v})$.
The encoding network is a residual network with 152 layers, ResNet152 \cite{he2016cvpr}, and is pre-trained on the ImageNet dataset \cite{russakovsky2015ijcv};
the rest of the network (the text-only part) is pre-trained on CANTAB.
We follow with a second pre-training step, by using the multi-modal FlickR-8K dataset.
Finally, as in the case of the text-only RNN, we use fine-tuning to perform domain adaptation.

\begin{table*}[!htbp]
    \centering
    \newcommand{\res}[2]{#1 \scriptsize{$\pm$ #2}}
    \newcommand{\ssr}[1]{\em{#1}}  
    \newcommand{\nnn}[2]{#1 \scriptsize{(#2\%)}}
    \newcommand{\ii}[1]{\scriptsize{\color{gray} #1}}
    \setlength{\tabcolsep}{3pt}
    \caption{%
        Word error rate (WER) on the proposed \uav dataset using adapted language models.
        Rows 1--4 correspond to text-only models, while rows 5--6 correspond to multi-modal models.
        We report the mean and two times the standard error across the five folds for each training size, $n$.
        A system using an unadapted language model obtains 56.2\% WER.
        Figures in italics indicate that the improvements from the best text-only model (row 4) to the multi-modal one (row 5) are statistically significant (based on McNemar's test at $p=0.05$).
    }
    \vspace{-0.25em}
    \begin{tabular}{clllrrrrrr}
        \toprule
        &              &                 & & \multicolumn{6}{c}{number of training sentences \scriptsize{(overlap with evaluation set)}} \\
                                             \cmidrule(l){5-10}
        & decoding     & rescoring       & img. assoc. & \nnn{2,048}{59}     & \nnn{4,096}{66}     & \nnn{8,192}{75}         & \nnn{16,384}{82}   & \nnn{32,768}{89}  & \nnn{65,536}{94} \\
        \midrule
        \ii{1} & FSG          & ---          &           & \res{26.22}{0.1} & \res{22.35}{0.3} & \res{19.02}{0.2} & \res{16.32}{0.1} & \res{14.52}{0.1} & \res{13.19}{0.1} \\
        \ii{2} & n-gram small & ---          &           & \res{15.91}{0.2} & \res{15.11}{0.1} & \res{14.65}{0.1} & \res{14.43}{0.1} & \res{14.18}{0.0} & \res{14.30}{0.1} \\
        \ii{3} & n-gram small & n-gram large &           & \res{15.27}{0.2} & \res{14.53}{0.3} & \res{13.67}{0.1} & \res{13.40}{0.1} & \res{12.98}{0.0} & \res{12.88}{0.2} \\
        \ii{4} & n-gram small & RNN          &           & \res{13.57}{0.1} & \res{12.43}{0.2} & \res{12.09}{0.1} & \res{11.89}{0.1} & \res{11.64}{0.1} & \res{11.48}{0.1} \\
        \midrule
        \ii{5} & n-gram small & RNN          & annotated & \res{13.43}{0.1} & \res{12.07}{0.2} & \ssr{\res{11.31}{0.1}} & \ssr{\res{11.00}{0.1}} & \ssr{\res{10.78}{0.1}} & \ssr{\res{10.73}{0.1}} \\
        \ii{6} & n-gram small & RNN          & generated & \res{13.49}{0.1} & \res{12.01}{0.2} &      \res{11.32}{0.1} &       \res{10.93}{0.2} &       \res{10.60}{0.1} &       \res{10.45}{0.0}  \\
        \bottomrule
    \end{tabular}
    \label{tab:adaptation-results}
    \vspace{-0.5em}
\end{table*}

\subsection{Auxiliary datasets}
\label{subsec:auxiliary-datasets}

In addition to the \uav dataset, we have also used the following public datasets:
\begin{description}
    \item[TED-LIUM] \cite{rousseau2014lrec}
        is a speech recognition corpus containing recordings of almost 1,500 TED talks (around 200 h).
        We used it to train the acoustic model.
    \item[CANTAB] \cite{williams2015arxiv}
        is a text corpus, collected from multiple sources, such as news or law.
        We use the data to initialize generic language models.
        The corpus contains around 14M sentences (252M words).
    \item[FlickR-8K] \cite{hodosh2013jair}
        is an image--text corpus: there are 8,000 images, each described by five captions (in total around 500K words).
        We use this corpus to initialize the multi-modal language model.
\end{description}

\section{Experimental results}
\label{sec:results}

\begin{figure}
    \centering
    \footnotesize
    \newcommand{\im}[1]{\includegraphics[width=3.0cm,height=2.25cm]{images/#1.jpg}}
    \newcommand{\gt}[1]{\textsf{gt:} \textit{#1}}
    \newcommand{\xenost}[1]{\textsf{txt:} \textit{#1}}
    \newcommand{\xenosi}[1]{\textsf{img:} \textit{#1}}
    \begin{tabular}[align=t]{cc}
        \im{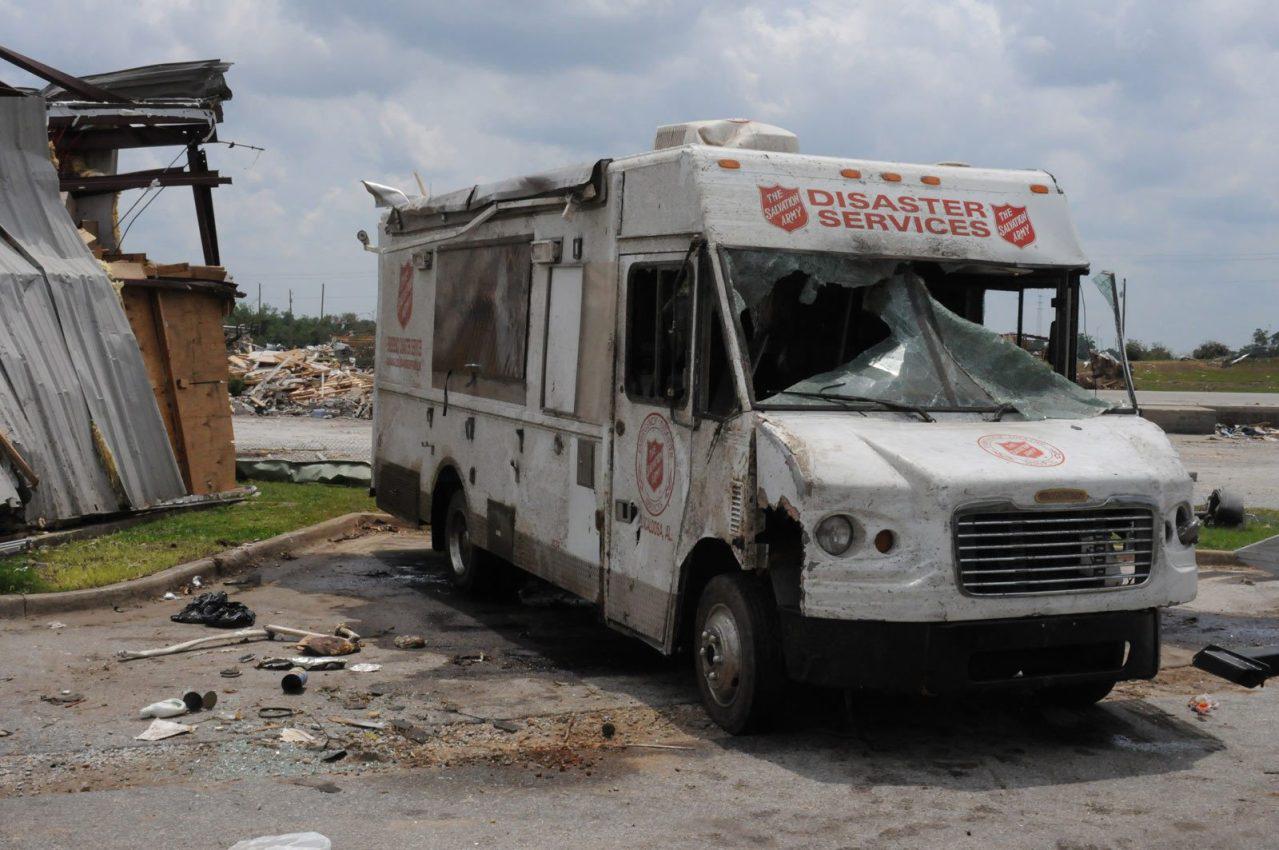} &
        \im{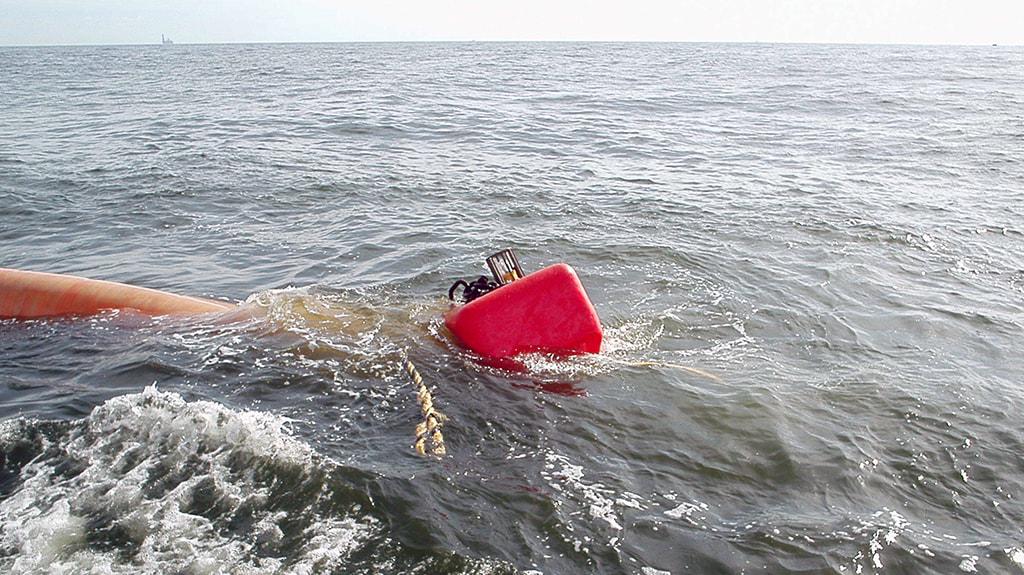} \\
        \gt{follow the truck} &
        \gt{film the buoy} \\
        \xenost{\color{BrickRed} follow the track} &
        \xenost{\color{BrickRed} film the boy} \\
        \xenosi{\color{PineGreen} follow the truck}  &
        \xenosi{\color{PineGreen} film the buoy} \\
        \midrule
        \im{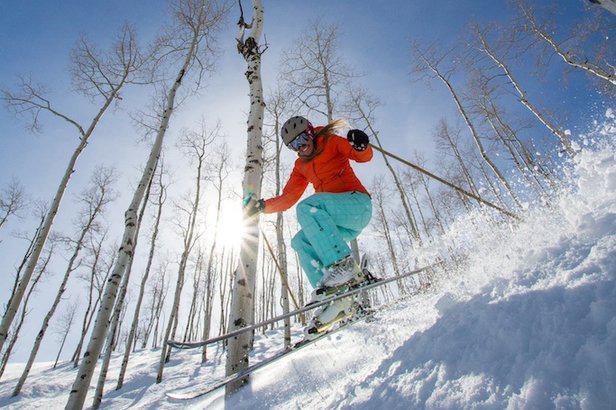} &
        \im{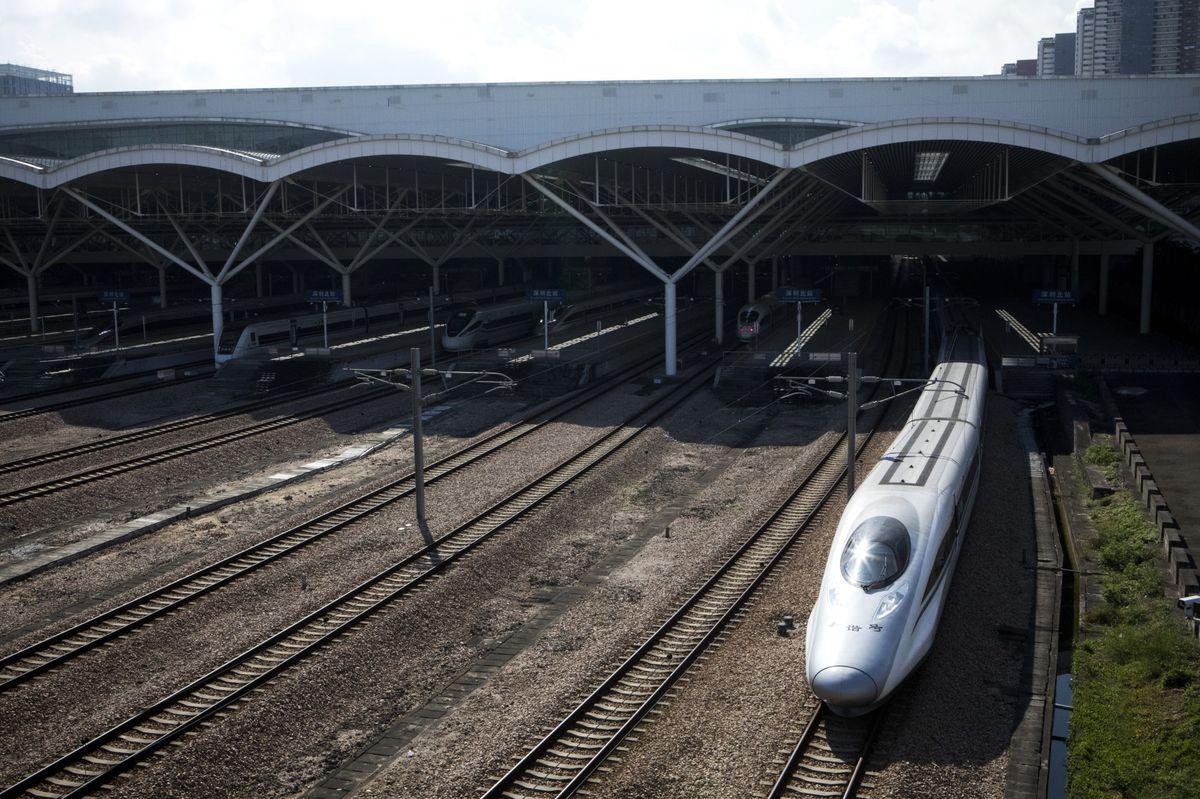} \\
        \gt{fly closer to the mountain} &
        \gt{fly to the railroad} \\
        \xenost{\color{PineGreen} fly closer to the mountain} &
        \xenost{\color{PineGreen} fly to the railroad} \\
        \xenosi{\color{BrickRed} fly closer to the man} &
        \xenosi{\color{BrickRed} fly to the train}  \\
    \end{tabular}
    \caption{
        Transcriptions of commands using the text-only RNN (\textsf{txt}) or the multi-modal RNN (\textsf{img}) language model.
        The groundtruth is denoted by \textsf{gt}.
        The first row shows success cases, while the last one shows failure cases.
    }
    \label{fig:qualitative-results}
\end{figure}

In this section we present the results on \uavtest dataset.

\textbf{Baseline systems.}
We compare the domain-adapted models against two baseline methods.
Both systems use the same acoustic model, which is trained on the TED-LIUM dataset,
but they differ in terms of the language model and the data used to train it:
the first system uses an n-gram trained on the CANTAB dataset and corresponds to a generic, unadapted speech recognition system;
the second system uses the gold finite state grammar (FSG), from which we sampled commands.
The unadapted system obtains a word error rate of 56.2\%,
while the system relying on the gold FSG obtains an WER of 11.7\%.
These results highlight the importance of adapting the language model to the target domain.

\textbf{Domain adaptation.}
The next experiment considers the case when we have access only to a partial list of commands at train time and we use those for domain adaptation.
The results for multiple language models and varying amounts of data are shown in table \ref{tab:adaptation-results};
in particular, the first part of the table, rows 1--4, presents the results for text-only adaptation.
As expected, the performance improves with more data and more flexible models.
The FSG is more reliant on data and it converges towards a good performance at a slower rate than the other models.
This behavior is expected, because if the exact command is missing from the training set, the FSG is unable to predict,
whereas the other models are more flexible and can interpolate missing words.
Rescoring improves the results further, with the RNN out-performing the n-gram model (rows 3 and 4).

\textbf{Multi-modal experiments.}
In the final experiment, we focus on language models that use visual information;
the corresponding results are in the second part of table \ref{tab:adaptation-results}, rows 5--6.
The two experiments differ in the images used for evaluation:
row 5 uses the standard \uavtest set of images collected manually from the internet;
row 6 uses an automatically selected set of images from ImageNet and FlickR-8K, similar to what we have done for the train set.
There are three main observations.
First, we notice that the visual information helps improve over the text-only model.
Second, the improvements are noticeable when we increase the data size, because the network is larger and needs more data to learn.
Third, having a different distribution at test time and possibly imperfect correspondences at train time,
does not impact the results: the differences between rows 5 and 6 are not statistical significant,
while we still obtain statistically significant improvements over the text-only model (row 4).

In figure \ref{fig:qualitative-results} we present qualitative results for the text-only (n-gram with RNN rescoring) and multi-modal models (n-gram with image RNN).
We show both success (first row) and failure cases (second row).
The multi-modal model is able to correct phonetically similar pairs which have visual grounding (\eg, buoy--boy, track--truck, trail--train),
but there are still cases where it biases too strongly towards the visual context (second row).

\section{Conclusions}

We have introduced a multi-modal dataset, \textsc{Kite}, for recognition of UAV commands.
Its evaluation part was manually annotated and curated, while the training part relied on more automatic approaches.
While the command--image associations used for training are likely to be imperfect, we have consistently found improvements over a text-only model.
This result confirms the benefits of the visual context for transcribing.
We have also shown the importance of adapting the language model and the benefits of using a more flexible model,
as its performance is less reliant on the quantity of data. 
Finally, we conclude with a couple of research directions that can be carried around our dataset:
\ia grounding the uttered commands in the images as a way of obtaining feedback from the system;
\ib improving the acoustic model by making it more robust to outdoor noises.

\noindent
\textbf{Acknowledgments.}
This research was partially supported by the POC-2015 P39-287 IAVPLN project.


\bibliographystyle{IEEEtran}
\bibliography{ref}

\end{document}